# Developing an AI-Based Psychometric System for Assessing Learning Difficulties and Adaptive System to Overcome: A Qualitative and Conceptual Framework (Ver 10th Mar 2024, this is a **working** paper.)

**Hu Yuxin**
University of Cambridge
yh552@cam.ac.uk

## Abstract

Learning difficulties pose significant challenges for students, impacting their academic performance and overall educational experience. These difficulties could sometimes put students into a downward spiral that lack of educational resources for personalized support consistently led to under-accommodation of students' special needs, and the student lose opportunities in the longer term academic and work development. This research aims to propose a conceptual framework for an adaptive AI-based virtual tutor system that incorporates psychometric assessment to support students with learning difficulties. This process involves the careful selection and integration of validated current mature psychometric scales that assess key dimensions of learning, such as cognitive abilities, learning styles, and academic skills. By incorporating scales that specifically assess these difficulties, the psychometric test will provide a comprehensive understanding of each student's unique learning profile and inform targeted interventions within the adaptive tutoring system. The paper also proposes using autoencoders to identify the latent patterns to generate the student's profile vector for collection of psychometric data, defining state space and action space representing the students' desired combination of images, sound and text engagements, employing extended Bayesian knowledge tracing and hierarchical model and Metropolis-Hastings to continuously estimate and monitor the student's performance in various psychometric constructs. The proposed system will leverage the capabilities of LLMs, visual generation models, and psychometric assessments to provide personalized instruction and support tailored to each student's unique learning characteristics and needs. By leveraging the capabilities of LLMs and visual generation models, the adaptive tutoring system will provide engaging and interactive learning experiences that cater to each student's individual needs and preferences.

I. Introduction

A. Background on learning difficulties and their impact on student performance

According to Sonali Nag & Margaret Snowling (2014), the prevalence and impact of specific learning difficulties and related psychiatric disorders on academic achievement have been a focal point of scholarly inquiry, demonstrating a significant variation influenced by the breadth of definitions employed. Studies indicate that specific learning difficulties affect 4-8% of the population under narrow definitions, with this figure rising to up to 18% when broader definitions are applied, inclusive of poor readers stemming from various underlying causes. Notably, the incidence of these difficulties peaks around the ages of 8 to 10 years. Epidemiological research conducted in diverse Indian locales underscores the interconnectedness between psychiatric disorders and educational underperformance. For instance, a study in Calicut found that 9.4% of children aged 8-12 were diagnosed with a psychiatric disorder, closely linked to school underachievement and reading/vocabulary challenges. Similarly, in Bangalore, 13% of children and adolescents aged 4-16 were identified with a psychiatric disorder, with a subset of up to 10% experiencing scholastic difficulties and an additional 2% grappling with both psychiatric disorders and academic underachievement.

Learning difficulties pose significant challenges for students, affecting their academic performance and overall educational experience. These difficulties encompass a wide range of cognitive, linguistic, behavioral, and emotional factors that can hinder students' ability to acquire, process, and apply knowledge effectively (American Psychiatric Association, 2013). Common learning difficulties include dyslexia, dyscalculia, and attention deficit hyperactivity disorder (ADHD), which can manifest in struggles with reading, writing, mathematics, and executive functioning (Fletcher et al., 2018). Students with learning difficulties often experience lower academic achievement, higher rates of grade retention, and increased risk of dropping out compared to their peers (Cortiella & Horowitz, 2014).

The global landscape reveals a prevalence rate of 10-20% for learning and developmental disabilities in high-income countries, with indications of even higher rates in low- and middle-income countries. Moreover, about 10% of primary school-aged children in low-income communities are affected by undetected vision problems that significantly hamper academic achievement, while hearing impairments also pose a considerable obstacle to educational attainment, notably affecting attention to auditory information and literacy development. These findings underscore the critical need for comprehensive assessments and interventions that address the multifaceted nature of learning difficulties and their impact on educational outcomes.

The reasons behind students' suboptimal learning are multifaceted, encompassing both individual and environmental factors. Individual factors such as motivation, self-regulation, and metacognitive skills play a crucial role in students' learning outcomes (Zimmerman, 2002). Students with learning difficulties may struggle with maintaining motivation, setting goals, and monitoring their progress, which can further exacerbate their academic challenges (Klassen, 2010). Environmental factors, including socioeconomic status, access to educational resources, and the quality of instruction, also contribute to students' learning experiences (Reardon, 2011). Students from disadvantaged backgrounds may face additional barriers to receiving appropriate support and accommodations for their learning needs (Shifrer et al., 2013).

Given the complex nature of learning difficulties and their impact on student performance, personalized support is essential in addressing these challenges. Traditional one-size-fits-all approaches to education often fail to meet the diverse needs of students with learning difficulties (Vaughn & Fuchs, 2003). Psychometric assessments have been widely used to identify individual learning characteristics and tailor educational interventions accordingly (Fletcher &

Vaughn, 2009). However, these assessments can be time-consuming, resource-intensive, and may not fully capture the dynamic nature of students' learning needs (Nag & Snowling, 2012).

However, the disparity in access to educational resources and personalized education remains a significant challenge in the United Kingdom and globally, exacerbating the educational divide between different socio-economic groups. In the UK, the Department for Education (2021) reported that approximately 15% of primary schools and 17% of secondary schools lack adequate resources to support effective teaching and learning. This situation is compounded in areas with high levels of poverty, where schools are often underfunded and understaffed, leading to a scarcity of personalized educational opportunities. For instance, a study by the Education Policy Institute (2020) found that students from disadvantaged backgrounds in England are, on average, 18.1 months behind their peers by the time they finish their GCSEs, highlighting the impact of resource limitations on educational outcomes.

Globally, the situation is even more dire, particularly in low- and middle-income countries. The United Nations Educational, Scientific and Cultural Organization (UNESCO, 2020) reports that over 260 million children and adolescents worldwide do not have access to schooling, and among those who do, a significant portion attends schools lacking basic resources such as textbooks, teaching materials, and trained teachers. This lack of resources severely hampers the delivery of personalized education, which is crucial for addressing individual learning needs and overcoming barriers to education. For example, the World Bank (2019) estimates that in Sub-Saharan Africa, only about 20% of schools have access to electricity, further limiting the potential for digital educational resources and personalized learning technologies that require power.

The scarcity of educational resources and personalized learning options is not solely a matter of physical resources but also reflects inadequacies in teacher training and support. According to the Organization for Economic Co-operation and Development (OECD, 2019), over 69% of teachers in OECD countries report a need for professional development in differentiated teaching practices, essential for personalized education. This gap is more pronounced in developing countries, where the UNESCO Institute for Statistics (UIS, 2018) notes that nearly 85% of teachers have not received sufficient training in inclusive teaching strategies, crucial for catering to diverse student needs.

The advent of artificial intelligence (AI) and advanced language models has opened up new possibilities for developing adaptive, personalized learning systems (Atari et al., 2022; Kuribayashi et al., 2023). Large Language Models (LLMs) and Visual Generation Models, such as GPT-4 and DALL-E, have demonstrated remarkable capabilities in understanding and generating human-like text and images (OpenAI, 2023). These AI technologies can be leveraged to create interactive, multi-modal learning experiences that adapt to students' individual characteristics and needs (Nag et al., 2017). By integrating psychometric assessments with AI-based tutoring systems, we can develop a more comprehensive and dynamic approach to supporting students with learning difficulties.

B. The potential of AI-based virtual tutoring systems for personalized learning

One of the key advantages of AI-based virtual tutoring systems is their ability to provide immediate, personalized feedback and support to students (Kulik & Fletcher, 2016). By analyzing students' responses, performance patterns, and engagement levels, these systems can dynamically adjust the content, difficulty, and pace of instruction to optimize learning outcomes (Nag et al., 2017). For example, an AI-based tutor can identify areas where a student is struggling and provide targeted explanations, hints, or additional practice exercises to address their specific challenges (Nye et al., 2014).

Moreover, AI-based virtual tutors can incorporate multi-modal learning materials, such as text, images, audio, and video, to enhance student engagement and understanding (Atari et al., 2022). LLMs like GPT-4 can generate human-like text explanations and prompts, while Visual Generation Models like DALL-E can create visually appealing and contextually relevant images to support learning (OpenAI, 2023). This multi-modal approach can cater to different learning preferences and help students build connections between abstract concepts and real-world applications (Mayer, 2014).

Another potential benefit of AI-based virtual tutoring systems is their scalability and accessibility. Traditional one-on-one tutoring can be costly and resource-intensive, limiting its availability to many students who could benefit from personalized support (VanLehn, 2011). In contrast, AI-based systems can be deployed at scale, providing on-demand tutoring to a large number of students simultaneously (Nag & Snowling, 2012). This can help bridge educational inequalities and ensure that all students, regardless of their socioeconomic background or geographic location, have access to high-quality, personalized learning support (Reardon, 2011).

However, the development and implementation of AI-based virtual tutoring systems also come with challenges and considerations. One critical aspect is the need for robust psychometric assessments to accurately identify students' learning characteristics, strengths, and weaknesses (Fletcher & Vaughn, 2009). These assessments should be integrated into the AI-based system to inform the initial personalization of learning content and to continuously monitor student progress (Kuribayashi et al., 2023). Collaborative efforts between psychometricians, educators, and AI experts are essential to ensure that these assessments are valid, reliable, and aligned with the learning objectives of the virtual tutoring system (Nag et al., 2017).

Another challenge is ensuring that AI-based virtual tutors are designed with ethical considerations in mind. As LLMs and Visual Generation Models can potentially produce biased or inconsistent outputs (OpenAI, 2023), it is crucial to develop rigorous testing and validation processes to mitigate the risk of perpetuating stereotypes or providing misleading information to students (Kuribayashi et al., 2023). Additionally, the privacy and security of student data must be prioritized, with clear guidelines on how this data is collected, stored, and used to personalize learning experiences (Nag & Snowling, 2012).

Despite these challenges, the potential of AI-based virtual tutoring systems for personalized learning is immense. By leveraging the capabilities of LLMs, Visual Generation Models, and psychometric assessments, these systems can provide adaptive, engaging, and effective learning support to students with diverse learning needs. As research in this area advances, it is essential to foster collaborative efforts among educators, psychometricians, and AI experts to develop and refine these systems, ensuring that they are grounded in sound pedagogical principles and aligned with ethical considerations (Atari et al., 2022).

C. The importance of psychometric assessment in adaptive tutoring

One of the primary goals of psychometric assessment in adaptive tutoring is to create a comprehensive understanding of each student's learning needs (Nag & Snowling, 2012). This involves assessing a wide range of cognitive, linguistic, and behavioral factors that contribute to learning difficulties, such as working memory, processing speed, and executive functioning (Fletcher et al., 2018). By gathering this information, adaptive tutoring systems can tailor their instructional strategies, content, and feedback to address students' specific challenges and optimize learning outcomes (Atari et al., 2022).

Psychometric assessments also enable the development of dynamic student profiles that can be continuously updated based on students' interactions with the adaptive tutoring system (Nag et al., 2017). As students engage with the system, their performance data, response patterns, and engagement levels can be analyzed to refine the understanding of their learning characteristics and adjust the personalization of instruction accordingly (Kuribayashi et al., 2023). This ongoing assessment process ensures that the adaptive tutoring system remains responsive to students' evolving needs and progress (OpenAI, 2023).

Moreover, psychometric assessments can help identify students who may be at risk of falling behind or experiencing persistent learning difficulties (Fletcher & Vaughn, 2009). By detecting these challenges early on, adaptive tutoring systems can provide targeted interventions and support to prevent the widening of achievement gaps (Reardon, 2011). This is particularly important for students from disadvantaged backgrounds, who may face additional barriers to accessing appropriate educational resources and support (Shifrer et al., 2013).

D. Research goals:

1. Examine Current Common Learning Difficulties and Reasons for Students' Suboptimal Learning: This goal aims to conduct a comprehensive analysis of prevalent learning difficulties affecting student performance. The focus will be on identifying the cognitive, linguistic, behavioral, and emotional factors that contribute to these difficulties.

2. Create a General Psychometric Test to Capture Individual Learning Characteristics: The integration of psychometric assessment into adaptive tutoring systems plays a pivotal role in discerning the unique learning characteristics of each student. By developing a general psychometric test that identifies key dimensions of learning, such as cognitive abilities and learning styles, educators can tailor their instructional approaches to match individual needs more effectively. This process involves the careful selection and integration of validated psychometric scales, along with meticulous considerations for the administration and interpretation of these tests to ensure accuracy and relevance. Traditional psychometric assessments, however, come with limitations, including their static nature and potential biases. Herein lies the potential of AI-based approaches, which can transcend these boundaries by continuously adapting and updating in response to the student's evolving learning profile. This dynamic method not only offers a more nuanced understanding of the learner but also introduces a transformative shift in personalized education, promising to significantly enhance the efficacy of adaptive tutoring systems.

3. Develop an AI-Based System That Adapts to Students' Needs: This goal focuses on the creation of an AI-based virtual tutoring system that leverages the insights gained from the psychometric assessments to provide personalized learning experiences. The system will use advanced algorithms and adaptive learning technologies to adjust the content, pace, and style of instruction based on real-time feedback and student performance data. This adaptive approach aims to address the specific challenges and leverage the strengths identified through the psychometric test, enhancing learning outcomes for students with difficulties.

II. Understanding Learning Difficulties

Below is a high level summarization of common learning difficulties:

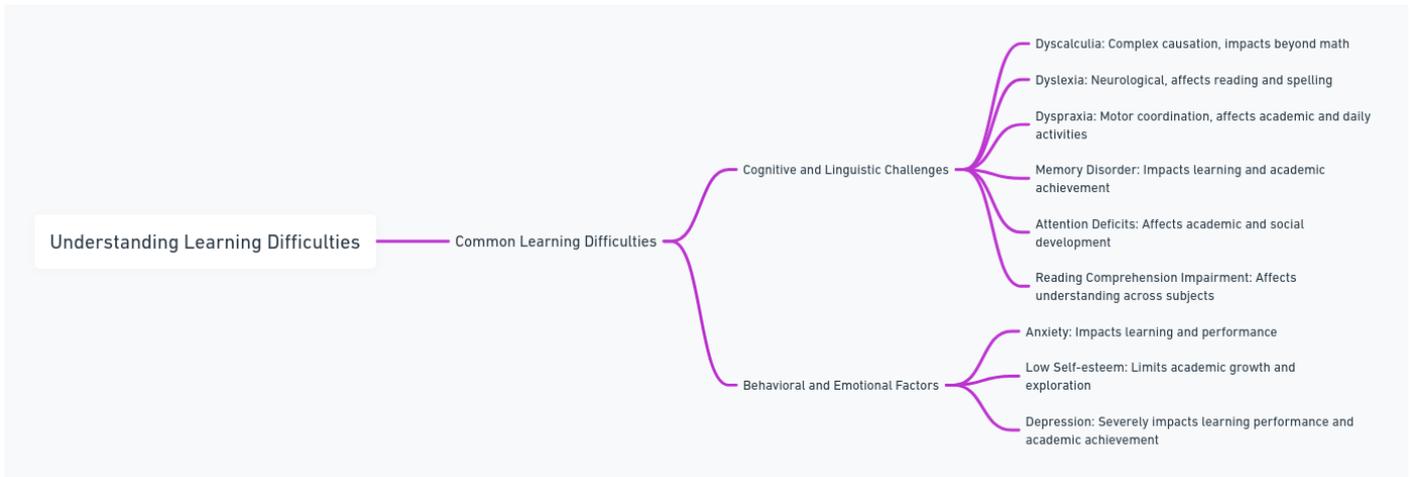

A. Common learning difficulties

1. Cognitive and linguistic challenges

1.1 Dyscalculia

Dyscalculia is identified as a specific learning disability that significantly hinders an individual's ability to understand numbers and learn mathematics facts. It's not just about struggling with math but rather a fundamental difficulty in processing numerical information, leading to challenges in performing calculations, understanding time, using money, and even playing games that require strategic quantification (American Psychiatric Association, 2013). The causation of dyscalculia is complex and multifaceted, involving genetic, neurobiological, and environmental components. Neuroimaging studies have highlighted abnormalities in the parietal lobe of individuals with dyscalculia, indicating a neurological basis for the disorder. This brain region is crucial for numerical cognition, including the processing of numerical magnitude and the execution of arithmetic operations (Butterworth, Varma, & Laurillard, 2011). Genetic studies also suggest a heritable component, with a higher prevalence of dyscalculia in families with a history of the disorder (Shalev, 2007).

The causation of dyscalculia is complex and multifaceted, involving genetic, neurobiological, and environmental components. Neuroimaging studies have highlighted abnormalities in the parietal lobe of individuals with dyscalculia, indicating a neurological basis for the disorder. This brain region is crucial for numerical cognition, including the processing of numerical magnitude and the execution of arithmetic operations (Butterworth, Varma, & Laurillard, 2011). Genetic studies also suggest a heritable component, with a higher prevalence of dyscalculia in families with a history of the disorder (Shalev, 2007).

The impact of dyscalculia on learning performance extends beyond mathematics. Students with dyscalculia often experience significant anxiety regarding math, which can affect their overall academic performance and self-esteem. The difficulty in understanding basic mathematical concepts and operations can hinder progress not only in math-related subjects but also in any academic area that requires quantitative reasoning. Additionally, the challenges associated with dyscalculia can lead to avoidance of math-related activities, further exacerbating the learning gap (Mazzocco & Myers, 2003). Moreover, the social implications of dyscalculia should not be underestimated. Struggles with everyday tasks that require numerical understanding, such as telling time or managing money, can lead to diminished life skills and potential challenges in adulthood, including difficulties with employment and managing personal finances (Ramirez, Chang, Maloney, Levine, & Beilock, 2016).

1.2 Dyslexia

Dyslexia is defined by the International Dyslexia Association as a specific learning disability that is neurological in origin, characterized by difficulties with accurate and/or fluent word recognition and by poor spelling and decoding abilities (International Dyslexia Association, 2021). These difficulties typically result from a deficit in the phonological component of language that is often unexpected in relation to other cognitive abilities and the provision of effective classroom instruction.

Research indicates that dyslexia has a strong genetic and biological basis, with studies showing that it tends to run in families (Peterson & Pennington, 2012). Neurobiological research has identified differences in the way the brain of a person with dyslexia processes language. Functional MRI studies have shown that individuals with dyslexia display reduced activity in areas of the brain typically involved in successful reading, such as the left temporoparietal cortex, during reading tasks (Shaywitz et al., 2002).

The impact of dyslexia on learning and academic performance can be profound and multifaceted. Students with dyslexia may experience significant challenges in reading fluency, comprehension, spelling, writing, and sometimes in oral language skills. These difficulties can hinder their ability to access and engage with the curriculum, leading to lower academic achievement and diminished educational opportunities (Snowling, 2000). Furthermore, the persistent challenges faced by students with dyslexia can affect their motivation, self-esteem, and overall mental health, often leading to increased rates of school dropout and fewer employment opportunities in adulthood (McNulty, 2003).

1.3 Dyspraxia

The American Psychiatric Association (2013) defines dyspraxia (DCD) in the Diagnostic and Statistical Manual of Mental Disorders (DSM-5) as a marked impairment in the development of motor coordination, significantly interfering with academic achievement or activities of daily living. This definition highlights the impact of dyspraxia on an individual's functioning beyond just the motor difficulties, encompassing academic and daily life challenges.

The etiology of dyspraxia is not fully understood, but it is believed to involve anomalies in brain development that affect the way the brain processes information, leading to difficulties in planning and coordinating physical movements. There is evidence to suggest that dyspraxia has a genetic component, as it often runs in families. Additionally, factors such as premature birth, low birth weight, and maternal alcohol or drug use during pregnancy have been associated with an increased risk of developing the condition (American Psychiatric Association, 2013; Kirby & Peters, 2019).

Dyspraxia can have a profound impact on a student's academic performance and school experience. The motor skills difficulties associated with dyspraxia can make it challenging for students to perform tasks such as handwriting, using scissors, or participating in physical education, which are often important components of the school curriculum. These challenges can lead to frustration, decreased motivation, and avoidance of activities that require motor coordination.

Moreover, dyspraxia can affect other areas of development that are crucial for learning, such as organization, time management, and the ability to follow multi-step instructions. Students with dyspraxia may also experience difficulties with speech and language, further complicating their ability to communicate effectively in the classroom setting (Barnhart et al., 2003; Missiuna et al., 2014).

The impact of dyspraxia on learning and academic performance underscores the importance of early identification, supportive interventions, and accommodations in the educational environment. For instance, allowing the use of typing instead of handwriting, providing additional time for tasks, and using visual aids can help mitigate some of the challenges faced by students with dyspraxia. Adaptive AI-based tutoring systems that can offer customized learning experiences and accommodate various learning styles present a promising approach to supporting students with dyspraxia in achieving their educational potential.

1.4 Memory Disorder

Memory disorders are characterized by significant impairment in the ability to remember information, events, or skills. This can include difficulties with short-term memory (the ability to hold information in mind for a short period), long-term memory (the ability to store information over time), and working memory (the ability to manipulate information in one's mind). Memory disorders can be standalone issues or symptoms of other neurological conditions, such as Alzheimer's disease, dementia, or brain injuries (Tulving & Craik, 2000).

The causes of memory disorders can vary widely, depending on the type of memory impairment and the individual's overall health condition. Some common causes include neurological diseases (such as Alzheimer's disease and other dementias), traumatic brain injury, stroke, and certain psychiatric conditions. Additionally, environmental factors, stress, and lack of sleep can temporarily affect memory functioning. Genetic predispositions also play a role in the susceptibility to memory disorders, especially in conditions like Alzheimer's disease (Nadel & Moscovitch, 1997; Schacter, 1996).

Memory disorders can severely affect an individual's learning performance and academic achievement. Difficulties in encoding and storing new information can make it challenging for students to learn new concepts or skills. Problems with retrieval can hinder their ability to recall previously learned material during tests or practical applications. This can lead to frustration, decreased motivation, and lower self-esteem, further exacerbating the learning difficulties.

For students with memory disorders, traditional teaching methods that rely heavily on rote memorization and recall of information may not be effective. Instead, teaching strategies that emphasize understanding, application, and repetition, along with the use of mnemonic devices and visual aids, can help improve learning outcomes. Adaptive AI-based tutoring systems that can personalize learning content and pace according to the student's memory capabilities offer a promising approach to supporting these students. Such systems can provide repeated exposure to material, use varied presentation methods, and incorporate memory aids to enhance learning and retention (Baddeley, 2013).

1.5 Attention Deficits

Attention deficits are characterized by an individual's persistent difficulty in maintaining focus on tasks that require sustained mental effort. This may include academic assignments, reading, or listening to lectures. Individuals with attention deficits might appear to listen when spoken to directly, but their attention can be easily diverted. They may also struggle with following through on instructions and fail to complete schoolwork, chores, or duties in the workplace due to attentional lapses.

The etiology of attention deficits is complex and multifaceted, involving genetic, environmental, and neurobiological factors. Research suggests that variations in genes related to dopamine regulation play a significant role, affecting neurotransmitter systems that are critical for attention and executive function. Environmental factors, such as exposure to toxins, premature birth, and low birth weight, have also been implicated. Neuroimaging studies have identified structural and functional differences in the brains of individuals with attention deficits, particularly in areas related to attention control, executive function, and motivation (Faraone et al., 2005; Shaw et al., 2007).

Attention deficits can significantly impede academic and social development. In educational settings, students with attention deficits may have difficulty organizing tasks, prioritizing homework, and adhering to deadlines. Their inattentiveness in class can lead to gaps in knowledge, misunderstanding of key concepts, and poor test performance. The fluctuating attention span also affects their ability to engage in lengthy reading or writing tasks, resulting in incomplete assignments and frustration.

Socially, attention deficits can strain peer relationships. Difficulty paying attention to social cues and maintaining focus in conversations can be perceived as disinterest or disrespect, leading to misunderstandings and conflicts. Moreover, the impulsivity that often accompanies attention deficits can result in interrupting others, intruding on conversations, or making hasty decisions without considering consequences.

Educational strategies that accommodate attention deficits include structured routines, breaking down tasks into manageable parts, using visual aids to maintain focus, and incorporating technology that provides interactive and engaging learning experiences. Tailored interventions, such as cognitive-behavioral strategies and executive function training, can also support individuals in managing their attention deficits more effectively.

1.6 Reading Comprehension Impairment

RCI is characterized by difficulties in making sense of words and sentences despite fluent word recognition skills. Children with RCI can often read text out loud with proper pronunciation but face challenges in summarizing content, making inferences, understanding narratives, and connecting ideas across paragraphs. Unlike dyslexia, which primarily affects the ability to decode words, RCI impacts higher-level cognitive skills necessary for extracting and constructing meaning from text.

The causes of Reading Comprehension Impairment are multifactorial and can include both cognitive and linguistic factors. Key components contributing to RCI include deficits in vocabulary knowledge, background knowledge, inferencing skills, working memory, and attention. These deficits hinder the child's ability to connect new information with existing knowledge, maintain focus on the text, and actively engage in the processes required for deep comprehension (Catts, Hogan, & Adlof, 2005; Oakhill, Cain, & Elbro, 2015).

The impact of RCI on academic performance can be profound. Reading comprehension is fundamental to learning across all subjects; difficulties in understanding text can therefore affect not only language arts but also science, mathematics, and social studies, where text comprehension plays a crucial role in learning and assessment. Children with RCI may struggle to follow written instructions, understand exam questions, and engage with educational materials, leading to lower academic achievement and diminished educational opportunities.

Moreover, RCI can affect motivation and self-esteem. Children who struggle to understand text may become frustrated, leading to disengagement from reading activities and schoolwork. This disengagement can further exacerbate learning difficulties and create a cycle of academic challenges and decreased self-confidence.

2. Behavioral and emotional factors

2.1 Anxiety

Anxiety in educational settings is often characterized by excessive worry about academic performance, fear of failure, and apprehension regarding evaluations or the social aspects of the school environment. This can manifest as test anxiety, social anxiety, or generalized anxiety about school activities. Students with high levels of anxiety may exhibit symptoms such as restlessness, fatigue, difficulty concentrating, irritability, and sleep disturbances, which can hinder their academic engagement and performance (American Psychiatric Association, 2013).

The causes of anxiety are multifaceted, involving genetic, environmental, and psychological factors. Genetic predispositions can influence an individual's susceptibility to anxiety, while environmental factors such as stressful life events, family dynamics, and school-related pressures can trigger or exacerbate anxious feelings. Psychological aspects, including coping skills, self-esteem, and cognitive styles, also play critical roles in the development and maintenance of anxiety (Craske & Stein, 2016).

Anxiety can negatively affect various aspects of learning and academic performance. The cognitive load imposed by anxiety can reduce working memory capacity, making it difficult for students to process and retain new information, solve problems, and understand complex instructions (Eysenck, Derakshan, Santos, & Calvo, 2007). Test anxiety, in particular, can impair exam performance, leading to lower grades that do not accurately reflect the student's knowledge or abilities.

Moreover, anxiety can lead to avoidance behaviors, where students may skip classes, avoid participation, or not complete assignments to evade anxiety-provoking situations. This avoidance can further impede learning and academic achievement. Social anxiety can also affect group work and classroom participation, limiting opportunities for learning and social development.

2.2 Low Self-esteem

Low self-esteem in educational settings is often manifested as a student's belief that they are less capable, valuable, or deserving than their peers. This perception can lead to a reluctance to participate in classroom activities, avoidance of challenges, and a tendency to give up easily when faced with difficulty. Students with low self-esteem may also exhibit heightened sensitivity to criticism and a tendency to interpret neutral or ambiguous feedback negatively.

The development of low self-esteem in students can be attributed to a complex interplay of factors. Learning difficulties themselves can significantly contribute to feelings of inadequacy, as repeated experiences of failure or struggle in academic tasks can lead students to internalize a sense of incompetence. Environmental factors, such as the educational climate, parental expectations, and peer relationships, also play crucial roles. Negative comparisons with siblings or classmates, lack of positive reinforcement, and punitive or overly critical feedback can all undermine a student's self-esteem.

The impact of low self-esteem on learning performance and academic achievement is significant. Students with low self-esteem often fear failure to such an extent that they avoid taking risks or engaging fully in learning opportunities, limiting their academic growth and exploration of interests. This avoidance can lead to a self-fulfilling prophecy, where students do not develop their abilities or achieve their potential, reinforcing their negative self-perceptions.

Furthermore, low self-esteem can affect motivation and persistence. Students may lack the resilience to persevere through challenges, leading to decreased effort and withdrawal from academic tasks. Social interactions can also be impacted, as students with low self-esteem might withdraw from group activities or not seek help when needed, further isolating themselves and exacerbating feelings of inadequacy.

2.3 Depression

According to the World Health Organization (2020), depression goes beyond occasional sadness or academic stress; it is a chronic condition that affects how a person feels, thinks, and handles daily activities. In the context of education, students with depression may exhibit a lack of motivation, decreased energy, difficulties concentrating, changes in appetite or sleep patterns, and feelings of worthlessness or excessive guilt. These symptoms can significantly impair their ability to engage with academic materials, participate in class, and maintain social relationships.

The causes of depression in students are multifactorial, involving a combination of genetic, biological, environmental, and psychological factors. Learning difficulties can exacerbate the risk of developing depression, as students may experience chronic stress, frustration, and low self-esteem stemming from academic challenges. Environmental stressors such as family dynamics, social isolation, and bullying can also contribute. Neurobiological factors, including imbalances in neurotransmitters like serotonin and dopamine, play a role in the onset and maintenance of depression.

As Wei et al. (2016) mentioned, depression can severely impact a student's learning performance and academic achievement. Cognitive symptoms such as difficulties with concentration, memory, and decision-making can hinder their ability to absorb new information, complete assignments, and prepare for exams. The lack of motivation and energy may lead to missed classes and a decrease in academic engagement. Additionally, the social withdrawal often associated with depression can limit students' participation in group work and extracurricular activities, further isolating them from supportive networks and learning opportunities. Moreover, depression can affect a student's future educational aspirations and career prospects. Persistent depressive symptoms can lead to a downward spiral of academic underachievement and reduced life satisfaction, emphasizing the need for early identification and support.

III. Psychometric Assessment in Adaptive Tutoring

A. The role of psychometric assessment in understanding individual learning characteristics before the AI era

Psychometric assessment plays a vital role in understanding individual learning characteristics, particularly for students with learning difficulties such as dyscalculia, dyslexia, dyspraxia, memory disorders, attention deficits, reading comprehension impairment, anxiety, low self-esteem, and depression (Fletcher et al., 2018). By employing a range of standardized tests and measures, psychometric assessments provide valuable insights into students' cognitive abilities, academic skills, and socio-emotional functioning, which can inform the development of personalized interventions and support within an adaptive tutoring system (Nag & Snowling, 2012).

One key area where psychometric assessment contributes to understanding individual learning characteristics is in the identification of specific learning difficulties. For example, standardized tests of mathematical ability and numerical processing can help diagnose dyscalculia, a specific learning disability that affects an individual's ability to acquire and apply mathematical skills (Butterworth, 2005). Similarly, assessments of phonological processing, rapid automatized naming, and word reading can aid in the identification of dyslexia, a learning difficulty characterized by challenges in accurate and fluent word recognition and spelling (Snowling, 2001). Psychometric tools that evaluate motor coordination and planning, such as the Movement Assessment Battery for Children (Henderson et al., 2007), can assist in identifying dyspraxia, a developmental coordination disorder that can impact learning and academic performance (Gibbs et al., 2007).

In addition to identifying specific learning difficulties, psychometric assessment plays a crucial role in understanding students' cognitive profiles, including strengths and weaknesses in various cognitive domains. Comprehensive cognitive assessments, such as the Wechsler Intelligence Scale for Children (WISC-V; Wechsler, 2014) or the Woodcock-Johnson Tests of Cognitive Abilities (WJ IV COG; Schrank et al., 2014), provide valuable information about a student's verbal comprehension, visual-spatial processing, fluid reasoning, working memory, and processing speed. These cognitive profiles can help explain individual differences in learning and inform the selection of appropriate instructional strategies and accommodations within an adaptive tutoring system (Kuribayashi et al., 2023).

Psychometric assessment also contributes to understanding students' academic skills and performance across various domains. Standardized achievement tests, such as the Wechsler Individual Achievement Test (WIAT-III; Wechsler, 2009) or the Kaufman Test of Educational Achievement (KTEA-3; Kaufman & Kaufman, 2014), provide detailed information about a student's reading, writing, and mathematics skills. This information can help identify areas of strength and weakness, as well as discrepancies between a student's cognitive abilities and academic performance, which may indicate the presence of a learning difficulty (Fletcher & Vaughn, 2009). Furthermore, specific assessments of reading comprehension, such as the Gray Oral Reading Tests (GORT-5; Wiederholt & Bryant, 2012) or the Woodcock Reading Mastery Tests (WRMT-III; Woodcock, 2011), can help identify students with reading comprehension impairment and inform targeted interventions within an adaptive tutoring system (Atari et al., 2022).

Moreover, psychometric assessment plays a vital role in understanding students' socio-emotional functioning and mental health, which can significantly impact learning and academic performance. Standardized measures of anxiety, such as the Screen for Child Anxiety Related Emotional Disorders (SCARED; Birmaher et al., 1999) or the Multidimensional Anxiety Scale for Children (MASC-2; March, 2013), can help identify students experiencing anxiety that may interfere with their learning. Similarly, assessments of self-esteem, such as the Rosenberg Self-Esteem Scale (Rosenberg, 1965) or the Coopersmith Self-Esteem Inventory (Coopersmith, 1981), can provide insights into students' self-perceptions and self-worth, which can influence their academic motivation and engagement

(OpenAI, 2023). Measures of depression, such as the Children's Depression Inventory (CDI-2; Kovacs, 2010) or the Beck Depression Inventory for Youth (BDI-Y; Beck et al., 2001), can help identify students experiencing depressive symptoms that may impact their learning and overall well-being.

The information gathered through psychometric assessment can be used to develop comprehensive student profiles that capture individual learning characteristics, strengths, and challenges. These profiles can then be integrated into an adaptive tutoring system that leverages the capabilities of Large Language Models (LLMs) and Visual Generation Models to provide personalized instruction, feedback, and support (Nag et al., 2017). For example, an adaptive tutoring system may use a student's cognitive profile to adjust the complexity and presentation of learning materials, or it may provide targeted interventions for specific learning difficulties, such as phonological awareness training for students with dyslexia (Kuribayashi et al., 2023).

However, it is essential to acknowledge the limitations and potential biases of psychometric assessment tools, as well as the need for ongoing monitoring and updating of student profiles within an adaptive tutoring system (Fletcher et al., 2018). Collaborative efforts among educators, psychometricians, and AI experts are necessary to ensure that psychometric assessments are valid, reliable, and culturally appropriate, and that the insights derived from these assessments are accurately integrated into the adaptive tutoring system (Atari et al., 2022).

B. Developing a general psychometric test

1. Identifying Constructs for Learning Difficulties

To create a comprehensive psychometric test, we must first identify the key constructs that contribute to learning difficulties. Drawing from the top-down approach mentioned in the paper "Evaluating General-Purpose AI with Psychometrics" (Wang et al., 2023), we can leverage existing theories and expert knowledge to determine these constructs. For example, the Cattell-Horn-Carroll (CHC) theory of cognitive abilities (Schneider & McGrew, 2018) provides a well-established framework for understanding various cognitive abilities, such as fluid reasoning, short-term memory, and processing speed, which are crucial for learning. Additionally, we can consider constructs related to academic skills, such as reading comprehension (Oakhill et al., 2015), mathematical problem-solving (Swanson et al., 2018), and written expression (Graham et al., 2016), as well as constructs related to motivation and self-regulation (Zimmerman, 2002).

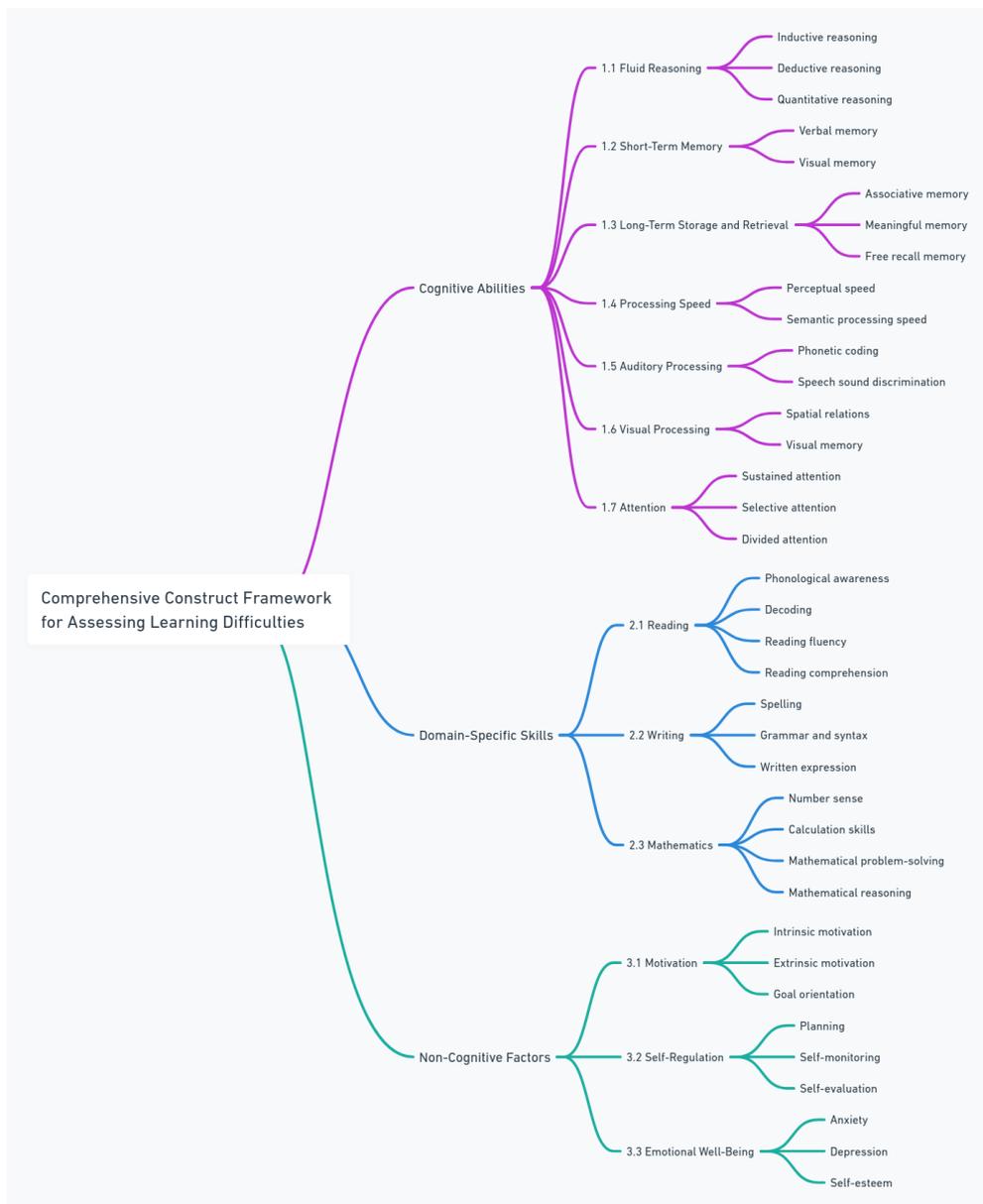

This construct framework is grounded in the Cattell-Horn-Carroll (CHC) theory of cognitive abilities (Schneider & McGrew, 2018), which provides a comprehensive taxonomy of cognitive abilities that are essential for learning. The framework includes key cognitive abilities such as fluid reasoning, short-term memory, processing speed, and attention, which have been consistently linked to academic performance (Finn et al., 2014; Peng et al., 2019).

In addition to cognitive abilities, the framework incorporates domain-specific skills in reading, writing, and mathematics. These skills are critical for academic success and are often the areas where students with learning difficulties struggle the most (Graham et al., 2016; Swanson et al., 2015). By assessing these skills, the AI-based psychometric system can identify specific areas of weakness and inform targeted interventions.

The framework also considers non-cognitive factors, such as motivation, self-regulation, and emotional well-being. These factors have been shown to significantly impact students' learning and academic performance (Roosta et al., 2021; Zimmerman, 2002). By including these factors in the construct, the AI-based psychometric system can provide a more comprehensive understanding of students' learning difficulties and guide interventions that address both cognitive and non-cognitive challenges.

To ensure the construct framework's compatibility with an AI-based psychometric system, the proposed constructs and sub-constructs have been carefully selected to be measurable through computer-based assessments. The framework focuses on abilities and skills that can be reliably assessed using text-based prompts, multiple-choice questions, and simple interactive tasks, which can be easily generated and adapted by Large Language Models (LLMs) and Visual Generation Models (Wang et al., 2023).

2. Operationalizing the constructs and sub-constructs into measurable indicators;

Building upon the previously identified constructs for learning difficulties in a computer-based assessment, the next step is to operationalize these constructs and sub-constructs into measurable indicators. This process involves defining specific, observable, and quantifiable variables that accurately represent the underlying constructs (Duckworth & Yeager, 2015). By establishing clear operational definitions and measurement strategies, we can ensure that the assessment effectively captures the intended constructs and provides reliable data for identifying learning difficulties.

Please refer to the table below for reference:

| Category | Subcategory | Measure | Description |
|---|---|---|---|
| **Verbal Working Memory** | Forward digit span | Number of correctly recalled digits in the presented order | |
| | Backward digit span | Number of correctly recalled digits in the reverse order | |
| **Visual-Spatial Working Memory** | Pattern recognition accuracy | Percentage of correctly identified patterns | |
| | Pattern reproduction accuracy | Percentage of accurately reproduced patterns | |
| **Processing Speed** | **Visual Processing Speed** | Visual matching speed | Number of correct matches made within a given time limit |
| | | Visual matching accuracy | Percentage of correctly matched |

| Category | Subcategory | Measure | Description |
| --- | --- | --- | --- |
| | | | items |
| | **Auditory Processing Speed** | Auditory discrimination speed | Response time for identifying target sounds |
| | | Auditory discrimination accuracy | Percentage of correctly identified target sounds |
| **Attention** | **Sustained Attention** | Visual continuous performance accuracy | Percentage of correctly identified target stimuli |
| | | Auditory continuous performance accuracy | Percentage of correctly identified target sounds |
| | **Selective Attention** | Visual selective attention accuracy | Percentage of correctly identified target stimuli among distractors |
| | | Auditory selective attention accuracy | Percentage of correctly identified target sounds among distractors |
| **Academic Skills** | **Reading** | Word recognition: Word identification accuracy | Percentage of correctly identified words |
| | | Word identification speed | Number of words correctly identified within a given time limit |
| | | Reading comprehension: Passage comprehension accuracy | Percentage of correctly answered multiple-choice questions |
| | **Writing** | Spelling: Spelling accuracy | Percentage of correctly spelled words |
| | | Sentence construction: Sentence construction accuracy | Percentage of correctly constructed sentences |
| | **Mathematics** | Basic arithmetic: Arithmetic accuracy | Percentage of correctly answered arithmetic questions |
| | | Mathematical reasoning: Word problem accuracy | Percentage of correctly answered word problems |

| Category | Subcategory | Measure | Description |
|---|---|---|---|
| **Non-Cognitive Factors** | **Academic Self-Efficacy** | Perceived competence in reading/writing/mathematics | Average score on a Likert scale for self-reported competence |
| | **Learning Motivation** | Intrinsic motivation | Average score on a Likert scale for self-reported intrinsic motivation |
| | | Extrinsic motivation | Average score on a Likert scale for self-reported extrinsic motivation |
| | **Test Anxiety** | Perceived test anxiety | Average score on a Likert scale for self-reported test anxiety |

These measurable indicators provide a clear framework for developing specific tasks and questionnaire items that assess the identified constructs. For example, to measure verbal working memory, the assessment can include a forward and backward digit span task, with the number of correctly recalled digits serving as the indicator of performance. Similarly, reading comprehension can be assessed through multiple-choice questions based on short passages, with the percentage of correctly answered questions indicating the student's level of comprehension.

For non-cognitive factors, such as academic self-efficacy and learning motivation, self-report questionnaires using Likert scales can be employed. The average scores on these scales will serve as indicators of the student's perceived competence and motivation in different academic domains.

It is important to note that the development of tasks and questionnaire items should be guided by established psychometric principles, such as reliability, validity, and fairness (AERA, APA, & NCME, 2014). This involves ensuring that the tasks and items: (i) Consistently measure the intended constructs; (ii) Accurately represent the constructs they purport to measure; (iii) Are free from bias and accessible to all students, regardless of their background or abilities.

Furthermore, the operationalization of constructs should be an iterative process, involving pilot studies and refinements based on student performance data and feedback. This process helps to optimize the assessment's effectiveness in identifying learning difficulties and informing personalized interventions.

3. Implementing the assessment items using LLMs and Visual Generation Models

To ensure that the psychometric test is both effective and engaging for students, we can leverage the capabilities of LLMs and Visual Generation Models to create interactive and visually appealing test items. As mentioned in the paper "Evaluating General-Purpose AI with Psychometrics" (Wang et al., 2023), these AI technologies can generate human-like text explanations, prompts, and contextually relevant images to support learning and assessment.

For example, to assess reading comprehension, we can use LLMs to generate age-appropriate passages and questions that adapt to the student's reading level. Visual Generation Models can create accompanying images that illustrate key concepts or events in the passage, enhancing student engagement and understanding (Koć-Januchta et al., 2017). Similarly, for assessing mathematical problem-solving, LLMs can generate word problems that are tailored to the student's skill level, while Visual Generation Models can create visual representations of the problems, such as graphs or diagrams, to support problem-solving strategies (Rau, 2017).

Furthermore, we can incorporate game-like elements and interactive features into the test items to maintain student interest and motivation. For instance, students can earn virtual rewards or badges for completing test items, and the difficulty of the items can adaptively increase based on their performance (Roosta et al., 2021). LLMs can generate personalized feedback and hints to guide students through challenging items, providing a supportive and engaging testing experience (Farnadi et al., 2014).

Leveraging the capabilities of Large Language Models (LLMs) and Visual Generation Models is a crucial step in creating an engaging and effective computer-based assessment for identifying learning difficulties. These AI technologies can generate dynamic, personalized, and visually appealing content that adapts to each student's needs and abilities, enhancing the assessment experience and providing valuable insights into their learning challenges (Wang et al., 2023).

1. Cognitive Abilities 1.1 Working Memory
- Verbal working memory:
    - LLMs can generate age-appropriate digit span tasks, adjusting the length and complexity of the sequences based on the student's performance.
    - Visual Generation Models can create engaging visual representations of the digits, such as colorful numbers or familiar objects, to support memory retention.
- Visual-spatial working memory:
    - Visual Generation Models can generate a variety of patterns with different levels of complexity for pattern recognition and reproduction tasks.

- LLMs can provide clear, concise instructions and feedback to guide students through the tasks. 1.2 Processing Speed
- Visual processing speed:
  - Visual Generation Models can create sets of images for visual matching tasks, ensuring that the images are age-appropriate and visually distinct.
  - LLMs can generate engaging prompts and instructions to maintain student motivation throughout the timed tasks.
- Auditory processing speed:
  - LLMs can generate auditory discrimination tasks with different target sounds and distractors, adapting the complexity based on the student's performance.
  - Visual Generation Models can provide visual cues to accompany the auditory stimuli, enhancing student engagement and understanding. 1.3 Attention
- Sustained attention:
  - Visual Generation Models can create engaging and age-appropriate visual stimuli for continuous performance tasks, such as images of animals or objects.
  - LLMs can generate feedback and encouragement to help students maintain focus throughout the tasks.
- Selective attention:
  - LLMs can generate age-appropriate auditory selective attention tasks, such as identifying target words in a story or conversation.
  - Visual Generation Models can create visual selective attention tasks with target stimuli and distractors that are visually distinct and engaging.
2. Academic Skills 2.1 Reading
- Word recognition:
  - LLMs can generate word identification tasks with age-appropriate words, adjusting the difficulty based on the student's performance.
  - Visual Generation Models can create visual representations of the words, such as images or animations, to support word recognition and engagement.
- Reading comprehension:
  - LLMs can generate reading passages and multiple-choice questions that are tailored to the student's reading level and interests (Koć-Januchta et al., 2017).
  - Visual Generation Models can create images that illustrate key concepts or events in the passages, enhancing student understanding and engagement. 2.2 Writing
- Spelling:

- LLMs can generate spelling tasks with age-appropriate words, providing immediate feedback and corrections for typed responses.
- Visual Generation Models can create engaging visual prompts for spelling tasks, such as images of objects or scenes related to the target words.
- Sentence construction:
    - LLMs can generate sentence construction tasks with varying levels of complexity, offering guided prompts and feedback for typed responses.
    - Visual Generation Models can create visual representations of the sentences, such as images or animations, to support understanding and engagement. 2.3 Mathematics
- Basic arithmetic:
    - LLMs can generate arithmetic questions that adapt to the student's skill level, providing step-by-step guidance and feedback for typed responses.
    - Visual Generation Models can create visual representations of the arithmetic problems, such as images of objects or number lines, to support problem-solving strategies.
- Mathematical reasoning:
    - LLMs can generate word problems that are tailored to the student's skill level and interests, offering personalized hints and explanations (Rau, 2017).
    - Visual Generation Models can create visual representations of the word problems, such as graphs or diagrams, to support problem-solving strategies and engagement.
3. Non-Cognitive Factors 3.1 Academic Self-Efficacy
- LLMs can generate self-report questionnaire items that assess perceived competence in reading, writing, and mathematics, using age-appropriate language and contexts.
- Visual Generation Models can create engaging visual scales, such as emoji or star ratings, to accompany the Likert scale questionnaire items. 3.2 Learning Motivation
- LLMs can generate self-report questionnaire items that assess intrinsic and extrinsic motivation, using age-appropriate language and examples.
- Visual Generation Models can create engaging visual prompts or scenarios to accompany the questionnaire items, enhancing student interest and understanding. 3.3 Test Anxiety
- LLMs can generate self-report questionnaire items that assess perceived test anxiety, using age-appropriate language and contexts.
- Visual Generation Models can create visually appealing and calming designs for the questionnaire interface, helping to reduce student anxiety during the assessment.

## 4. Scoring and Interpreting Test Results

The integration of advanced psychometric techniques, notably Item Response Theory (IRT), into the computer-based assessment of learning difficulties is paramount for the precise measurement and meaningful interpretation of test results. IRT stands out as a sophisticated framework that models the intricate relationship between a student's latent abilities and their performance on specific test items. This approach facilitates a more accurate and insightful assessment of the constructs of interest, as highlighted by Embretson & Reise (2013) and Wang et al. (2023). Through the application of IRT models, such as the Rasch model or the two-parameter logistic model (2PL), educators can simultaneously estimate a student's latent abilities across various constructs. This method considers the difficulty and discrimination of each test item, offering a nuanced representation of a student's strengths and weaknesses beyond traditional scoring methods (De Ayala, 2013).

Utilizing IRT models to analyze assessment data yields estimates of a student's latent abilities in cognitive, academic, and non-cognitive domains. These estimates are typically expressed as theta ($\theta$) values on a standardized scale, providing a quantitative measure of the student's standing on each construct. Such detailed quantification allows for comparisons among students and the monitoring of individual progress over time. Furthermore, the derived latent abilities enable the creation of comprehensive student profiles, illustrating strengths and weaknesses across assessed constructs. These profiles, effectively communicated through visual aids like graphs or charts, enhance the understanding and dissemination of results to students, parents, and educators (Farnadi et al., 2014).

Adaptive testing and dynamic assessment methods, underpinned by IRT, further refine the evaluation process. Techniques such as the maximum information method or Bayesian sequential estimation method dynamically adjust the selection of test items based on the student's responses. This adaptive approach ensures that each item presented is optimally challenging for the student, enhancing the efficiency and precision of the assessment while maintaining engagement (van der Linden & Glas, 2010; Mukherjee et al., 2017).

The interpretation and communication of assessment results are critical to their utility. Narrative reports, which translate quantitative findings into accessible language and offer specific recommendations, play a vital role in making the results actionable. Tailoring these reports to the audience ensures that students, educators, and professionals receive relevant and understandable information, fostering a growth mindset and encouraging engagement with the learning process (Aschbacher & Herman, 1991; Dweck, 2006).

To support effective collaboration among stakeholders, the assessment outcomes and interpretations should be accessible through a user-friendly dashboard or platform. This system should not only facilitate secure information sharing but also provide resources for implementing personalized interventions and tracking progress, advocating a responsive, data-driven approach to supporting students with learning difficulties (De Ayala, 2013).

Continuous validation and refinement of the assessment tool are essential to maintain its accuracy, reliability, and fairness. This involves regular data analysis to evaluate the psychometric properties of test items and the interrelations among constructs, as well as incorporating feedback from users to enhance the assessment's relevance and effectiveness (Embretson & Reise, 2013).

5. Validation and Iterative Refinement

To ensure the reliability, validity, and effectiveness of the AI-based psychometric system for assessing learning difficulties, it is crucial to engage in a rigorous validation process and iterative refinement. This involves collecting data from a diverse sample of students, examining the psychometric properties of the test items, and leveraging the capabilities of Large Language Models (LLMs) and Visual Generation Models to refine the assessment based on student performance and feedback (Embretson & Reise, 2013; Gierl & Haladyna, 2012).

Validation is a critical step in establishing the credibility and trustworthiness of the assessment system. By collecting data from a representative sample of students, we can examine the difficulty, discrimination, and fit of the test items to the Item Response Theory (IRT) model (Embretson & Reise, 2013). This analysis will help identify items that may be too easy, too difficult, or fail to effectively discriminate between students with different ability levels. Additionally, assessing the test's convergent and discriminant validity by comparing its results with other established measures of cognitive abilities and academic skills will provide further evidence of the system's validity (Mukherjee et al., 2017).

Iterative refinement is another essential aspect of ensuring the assessment system's effectiveness and responsiveness to students' needs. By leveraging the capabilities of LLMs and Visual Generation Models, we can continuously improve the test items based on student performance and feedback (Gierl & Haladyna, 2012). For example, if certain items are found to be problematic during the validation process, LLMs can be used to generate new items that better target the desired difficulty and discrimination parameters. This approach allows for the dynamic adaptation of the assessment system to optimize its measurement properties and maintain its relevance over time.

Moreover, incorporating student feedback is a valuable source of information for refining the assessment system. If students report that certain items are confusing, unclear, or lack sufficient context, LLMs can be employed to revise the item wording, making it more precise and easily understandable (Wang et al., 2023). Similarly, Visual Generation Models can be used to create more informative and engaging visual aids that support students' comprehension and reduce cognitive load (Koć-Januchta et al., 2017).

The iterative refinement process should also involve collaboration with educational experts, psychometricians, and AI specialists to ensure that the refinements align with best practices in assessment design and ethical AI development (Luckin et al., 2016). This interdisciplinary approach will help maintain the system's technical robustness, pedagogical soundness, and adherence to ethical principles, such as fairness, transparency, and accountability (Danaher et al., 2017).

To support the validation and iterative refinement process, it is important to establish a comprehensive data collection and monitoring system that captures students' performance, interactions, and feedback in real-time (Kulik & Fletcher, 2016). This data can be used to continuously evaluate the assessment system's effectiveness, identify areas for improvement, and inform data-driven decision-making regarding refinements and adaptations.

Furthermore, it is essential to engage in ongoing validation studies to ensure that the assessment system remains reliable, valid, and fair over time, particularly as the student population and educational contexts evolve (Embretson & Reise, 2013). Regular revalidation can help detect and address any emerging biases, limitations, or unintended consequences of the system, ensuring that it continues to serve its intended purpose of supporting students with learning difficulties.

IV. AI-based Adaptive Tutoring System

A. Overview of the proposed AI-based virtual tutor system

The proposed AI-based virtual tutor system is designed to address the diverse learning needs of students by leveraging the capabilities of advanced AI technologies, such as Large Language Models (LLMs) and Visual Generation Models. This system aims to provide a personalized, engaging, and effective learning experience that adapts to each student's unique learning characteristics, as identified through an initial psychometric assessment and ongoing monitoring (Wang et al., 2023). The virtual tutor system is envisioned as a multi-modal learning platform that can generate and present educational content in various forms, including text, images, diagrams, and sound. By adjusting the content and the relative proportions of these media representations, the system can cater to the individual learning preferences and needs of each student (Mayer, 2014). For example, students with a stronger visual learning style may benefit from a higher proportion of images and diagrams, while those with auditory preferences may require more audio-based explanations (Koć-Januchta et al., 2017).

At the core of the AI-based virtual tutor system is a comprehensive student profiling mechanism that integrates data from the initial psychometric assessment and continuous monitoring of student interactions and performance (Kulik & Fletcher, 2016). This dynamic student profile serves as the foundation for the system's adaptive tutoring strategies, enabling it to tailor the content, presentation, and support to the evolving needs of each learner (Nag et al., 2017). The system's interactive communication capabilities, powered by natural language processing (NLP) and AI-driven dialogue management, allow for personalized feedback, guidance, and support (Nye et al., 2014). By engaging in natural conversations with students, the virtual tutor can understand their questions, concerns, and emotional states, providing timely and relevant responses that foster a supportive and motivating learning environment (Graesser et al., 2005).

As students interact with the AI-based virtual tutor system, their progress, engagement, and performance data are continuously tracked and analyzed. This information is used to update the student profiles dynamically, ensuring that the system maintains an accurate understanding of each learner's strengths, weaknesses, and evolving needs (Atari et al., 2022). Based on these updates, the virtual tutor can adapt its tutoring strategies in real-time, optimizing the learning experience and outcomes for each student (Kuribayashi et al., 2023). The proposed AI-based virtual tutor system represents a significant advancement in personalized, adaptive learning support for students with diverse learning needs. By harnessing the power of LLMs, Visual Generation Models, and advanced psychometric techniques, this system aims to bridge the gap between traditional one-size-fits-all education and truly individualized learning experiences (Luckin et al., 2016). As such, it holds the potential to revolutionize the way we address learning difficulties and support the academic success of all students.

B. Initial psychometric assessment and student profiling

The initial psychometric assessment and student profiling form the cornerstone of the AI-based virtual tutor system, enabling it to understand and adapt to each student's unique learning characteristics from the outset. To ensure a comprehensive, engaging, and age-appropriate assessment experience, we propose a 20-minute game-based test that incorporates the cognitive, academic, and non-cognitive factors identified in the operationalization framework (Wang et al., 2023). This approach aligns with the growing recognition of game-based assessments as a valid, reliable, and engaging means of measuring complex constructs in educational settings (Shute & Ventura, 2013; Mislevy et al., 2014).

The game-based assessment will be meticulously designed to measure the key constructs and sub-constructs outlined in the operationalization table, including working memory, processing speed, attention, reading, writing, mathematics, academic self-efficacy, learning motivation, and test anxiety. Each construct will be assessed through a series of short, interactive mini-games that are seamlessly integrated into an overarching narrative or theme, creating a cohesive and immersive assessment experience (Shute & Wang, 2016). The narrative or theme will be carefully selected to appeal to the target age group, ensuring that the assessment is both engaging and developmentally appropriate.

To assess verbal working memory, the game may present a story-based task where the student must remember and recall a sequence of characters or objects in the correct order. The difficulty of the task will adapt based on the student's performance, with the length and complexity of the sequences increasing as the student demonstrates mastery. Visual-spatial working memory can be measured through a puzzle-like mini-game that requires the student to recognize and reproduce patterns within the game environment. The patterns will vary in complexity and be presented through visually appealing and age-appropriate graphics generated by Visual Generation Models (Klingler et al., 2017).

Processing speed and attention will be assessed through timed challenges that involve identifying target stimuli among distractors, with the game adapting the difficulty level based on the student's performance. For example, a visual processing speed task may require the student to quickly match pairs of images, while an auditory processing speed task could involve identifying target sounds in a sequence. Sustained attention can be measured through continuous performance tasks, such as monitoring a virtual environment for specific events or characters, while selective attention can be assessed by having the student focus on relevant stimuli while ignoring distractors (Klingler et al., 2017).

Academic skills, such as reading, writing, and mathematics, will be evaluated through context-embedded tasks that simulate real-world applications. For instance, a reading comprehension mini-game may involve the student exploring a virtual library, interacting with characters, and answering questions about the information they encounter. The passages and questions will be generated by LLMs, ensuring that they are age-appropriate and align with the student's reading level (Shute et al., 2016). Writing skills can be assessed through a creative storytelling task, where the student must construct sentences or short passages based on visual prompts provided by Visual Generation Models. The complexity of the writing tasks will be adjusted based on the student's performance, with LLMs providing immediate feedback and guidance (Shute & Ventura, 2013).

Mathematical abilities will be measured through problem-solving scenarios that require the application of arithmetic and reasoning skills within the game world. For example, the student may need to solve puzzles or complete transactions using basic arithmetic operations, with the difficulty of the problems adapting to the student's skill level. LLMs will generate age-appropriate word problems that are embedded within the game narrative, while Visual Generation Models will create visual representations, such as graphs or diagrams, to support problem-solving strategies (Rau, 2017).

Non-cognitive factors, such as academic self-efficacy, learning motivation, and test anxiety, will be assessed through a combination of self-report items and behavioral indicators embedded within the game. For instance, the game may present the student with choices that reflect their perceived competence or ask them to rate their confidence in completing certain tasks using visually appealing scales or emoticons generated by Visual Generation Models. Learning motivation can be inferred from the student's persistence and engagement in the face of challenges, while test anxiety may be gauged through the student's response patterns and decision-making under pressure (Dicerbo, 2014; Ventura & Shute, 2013).

Throughout the game-based assessment, the AI-based virtual tutor system will collect and analyze data on the student's performance, interactions, and responses. This data will be used to generate a comprehensive student profile that captures the student's strengths, weaknesses, and learning preferences across the assessed constructs (Mislevy et al., 2014). The student profile will serve as the basis for the initial personalization of the tutoring system, informing the selection and presentation of learning content, the balance of media representations, and the nature of the system's interactive support (Kulik & Fletcher, 2016).

To ensure the validity and reliability of the game-based assessment, the system will employ advanced psychometric techniques, such as Item Response Theory (IRT) and Evidence-Centered Design (ECD), to calibrate the difficulty and discrimination of the assessment tasks and to model the relationship between observable performance and the underlying constructs of interest (Mislevy et al., 2014; Shute & Ventura, 2013). The assessment will also undergo rigorous validation studies to establish its psychometric properties and to refine its design based on empirical data and student feedback (Klingler et al., 2017).

C. Multi-modal knowledge representation and delivery

1. Adapting content presentation based on student characteristics

To effectively adapt the content presentation in the AI-based virtual tutor system, we need to consider a wide range of psychometric factors identified in the initial assessment and student profiling. These factors include cognitive abilities (working memory, processing speed, attention), academic skills (reading, writing, mathematics), and non-cognitive factors (academic self-efficacy, learning motivation, test anxiety). By incorporating these factors into the adaptation process, the system can provide a truly personalized learning experience that caters to each student's unique needs and preferences (Kulik & Fletcher, 2016).

The adaptation process begins by representing the student's profile as a high-dimensional vector that captures the various psychometric factors. Let's denote this vector as s = [θ_1, θ_2, ..., θ_n], where θ_i represents the student's ability or trait in the i-th psychometric factor. To reduce the dimensionality of this vector and identify latent patterns, we can apply techniques such as principal component analysis (PCA) or autoencoder neural networks (Goodfellow et al., 2016).

Next, we need to define a mapping function f that takes the student's profile vector s as input and outputs a set of content presentation parameters p = [p_1, p_2, ..., p_m]. These parameters can include text complexity, chunk size, the proportions of images, sound, and text, and other relevant factors. The mapping function can be learned from a dataset of student profiles and their corresponding optimal content presentation parameters using supervised learning techniques, such as neural networks or decision trees (Hastie et al., 2009).

Let's consider a simple example of adapting content presentation based on a student's verbal working memory and reading comprehension skills. Suppose the student's profile indicates a low verbal working memory capacity ($\theta_{vwm}$) and average reading comprehension ability ($\theta_{rc}$). The system can use these values to determine the optimal text complexity and chunk size for the student.

We can model the relationship between the student's abilities and the text complexity using a linear function:

$$text\_complexity = \alpha * \theta_{vwm} + \beta * \theta_{rc} + \gamma$$

where α and β are weights that determine the relative importance of verbal working memory and reading comprehension, respectively, and γ is a constant term. The values of α, β, and γ can be learned from a dataset of student profiles and their corresponding optimal text complexities using linear regression.

Similarly, the chunk size (i.e., the amount of information presented at a time) can be determined based on the student's verbal working memory capacity:

$$chunk\_size = \delta * \theta_{vwm} + \varepsilon$$

where δ is a weight and ε is a constant term, both of which can be learned from data.

By applying these functions, the system can dynamically adjust the text complexity and chunk size based on the student's profile. For example, if θ_vwm = -0.5 and θ_rc = 0.2, with learned parameters α = -0.8, β = 0.6, γ = 0.4, δ = 0.7, and ε = 3, the system would present text with a complexity of:

$$text\_complexity = -0.8 * -0.5 + 0.6 * 0.2 + 0.4 = 0.92$$

and a chunk size of:

$$chunk\_size = 0.7 * -0.5 + 3 = 2.65 \approx 3$$

This adaptation ensures that the student receives content that is neither too simple nor too challenging, optimizing their learning experience.

Let's also consider a simple neural network architecture for learning the mapping function f

```python
import numpy as np
import tensorflow as tf

# Define the input and output dimensions
input_dim = len(psychometric_factors)
output_dim = len(presentation_parameters)

# Define the neural network architecture
model = tf.keras.Sequential([
    tf.keras.layers.Dense(64, activation='relu', input_shape=(input_dim,)),
    tf.keras.layers.Dense(32, activation='relu'),
    tf.keras.layers.Dense(output_dim)
])

# Compile the model
model.compile(optimizer='adam', loss='mse')

# Train the model on a dataset of student profiles and optimal presentation parame
model.fit(student_profiles, optimal_parameters, epochs=100, batch_size=32)
```

In this example, the neural network consists of an input layer with dimensions equal to the number of psychometric factors, two hidden layers with 64 and 32 neurons, respectively, and an output layer with dimensions equal to the number of presentation parameters. The model is trained using the mean squared error (MSE) loss function and the Adam optimizer (Kingma & Ba, 2015).

Once the mapping function f is learned, the system can adapt the content presentation for a given student by following these steps:

1. Obtain the student's profile vector s from the initial assessment and ongoing interactions.
2. Apply the mapping function f to the profile vector s to obtain the content presentation parameters p.
3. Adjust the learning content based on the presentation parameters p, such as selecting appropriate text complexity, chunk size, and media proportions.
4. Present the adapted content to the student and monitor their learning progress and engagement.
5. Update the student's profile vector s based on their performance and feedback, and repeat the adaptation process.

To further enhance the adaptation process, we can incorporate reinforcement learning techniques, such as multi-armed bandits or contextual bandits (Lattimore & Szepesvári, 2020). These techniques allow the system to explore different content presentation strategies and learn the most effective ones for each student based on their feedback and performance. The system can balance exploration (trying new strategies) and exploitation (using the best-known strategies) to continuously improve the adaptation process.

For example, let's consider a simple multi-armed bandit algorithm called Upper Confidence Bound (UCB) (Auer et al., 2002):

```python
import numpy as np

# Define the number of content presentation strategies
num_strategies = len(presentation_strategies)

# Initialize the counts and rewards for each strategy
counts = np.zeros(num_strategies)
rewards = np.zeros(num_strategies)

# Define the exploration-exploitation trade-off parameter
c = 2

# Run the UCB algorithm for each student interaction
for t in range(num_interactions):
    # Select the strategy with the highest UCB score
    ucb_scores = rewards / (counts + 1e-5) + c * np.sqrt(np.log(t+1) / (counts +
    strategy = np.argmax(ucb_scores)

    # Apply the selected strategy and observe the reward
    reward = apply_strategy(strategy, student_profile)

    # Update the counts and rewards for the selected strategy
    counts[strategy] += 1
    rewards[strategy] += reward
```

In this example, the UCB algorithm maintains a count and a reward for each content presentation strategy. At each student interaction, it selects the strategy with the highest UCB score, which balances the exploitation of strategies with high average rewards and the exploration of strategies with low counts. The selected strategy is applied to the student, and the observed reward is used to update the counts and rewards. Over time, the algorithm learns the most effective strategies for each student based on their profile and feedback.

By combining supervised learning techniques for mapping student profiles to content presentation parameters and reinforcement learning techniques for exploring and adapting presentation strategies, the AI-based virtual tutor system can provide a highly personalized and effective learning experience that considers a wide range of psychometric factors.

2. Optimizing the use of images, sound, and text

To optimize the use of images, sound, and text in the AI-based virtual tutor system, we can employ a multi-modal reinforcement learning approach (Wang et al., 2020). The goal is to learn an optimal policy that selects the most effective combination of media representations for each student based on their profile and learning progress.

The reinforcement learning framework consists of the following components:

- State space (S): Represents the student's current profile and learning progress, encoded as a vector of features (e.g., cognitive abilities, academic skills, engagement levels, and topic mastery).
- Action space (A): Represents the possible combinations of images, sound, and text that the system can present to the student. Each action is a vector that specifies the proportions of each media type (e.g., [0.4, 0.3, 0.3] for 40% images, 30% sound, and 30% text).
- Reward function (R): Measures the effectiveness of the selected media combination based on the student's learning outcomes (e.g., improvement in assessment scores, engagement, or retention). The reward can be calculated as a weighted sum of these metrics, with weights determined by the system's objectives.
- Transition function (T): Models the probability of transitioning from one state to another based on the selected action and the student's profile. The transition function can be learned from data using techniques like dynamic Bayesian networks (DBNs) (Ghahramani, 2001).

The system learns the optimal policy π*(s) that maps states to actions by maximizing the expected cumulative reward over time:

$$\pi*(s) = \mathrm{argmax}_a Q*(s, a)$$

where Q*(s, a) is the optimal action-value function, which represents the expected cumulative reward of taking action a in state s and following the optimal policy thereafter. The Q-function can be learned using algorithms like Q-learning or SARSA (Sutton & Barto, 2018).

During the learning process, the system interacts with the student and updates the Q-function based on the observed rewards and state transitions. The Q-function update rule for Q-learning is:

$$Q(s,a) \leftarrow Q(s,a) + \alpha * [R(s,a) + \gamma * max_{a'} Q(s',a') - Q(s,a)]$$

where α is the learning rate, γ is the discount factor, s' is the next state, and a' is the next action.

To balance exploration and exploitation, the system can use techniques like ε-greedy exploration, where it selects a random action with probability ε and the greedy action $argmax_a Q*(s,a)$ with probability 1-ε.

Over time, the system learns to adapt the proportions of images, sound, and text based on the student's profile and learning progress, optimizing the multi-modal knowledge representation and delivery.

Here's a simple example of how the Q-learning algorithm can be implemented in Python:

```python
import numpy as np

# Define the state and action spaces
state_space = np.array([[0, 0], [0, 1], [1, 0], [1, 1]])
action_space = np.array([[0.2, 0.3, 0.5], [0.4, 0.4, 0.2], [0.6, 0.2, 0.2]])

# Define the reward function
def reward(state, action):
    # Placeholder reward function
    return np.random.randint(0, 10)

# Define the Q-learning parameters
alpha = 0.1
gamma = 0.9
epsilon = 0.1

# Initialize the Q-function
Q = np.zeros((len(state_space), len(action_space)))

# Run the Q-learning algorithm
for episode in range(1000):
    state = np.random.choice(len(state_space))
    
    while True:
        # Select an action using ε-greedy exploration
```

```
26.         if np.random.rand() < epsilon:
27.             action = np.random.choice(len(action_space))
28.         else:
29.             action = np.argmax(Q[state])
30.
31.         # Take the action and observe the reward and next state
32.         next_state = np.random.choice(len(state_space))
33.         r = reward(state_space[state], action_space[action])
34.
35.         # Update the Q-function
36.         Q[state, action] += alpha * (r + gamma * np.max(Q[next_state]) - Q[state, action])
37.
38.         # Transition to the next state
39.         state = next_state
40.
41.         # Check if the episode is done
42.         if np.random.rand() < 0.1:
43.             break
44.
45. # Print the learned Q-function
46. print(Q)
```

In this example, the state space represents different student profiles (e.g., [0, 0] for low verbal working memory and low reading comprehension), and the action space represents different combinations of images, sound, and text. The reward function is a placeholder that returns a random reward for demonstration purposes. The Q-learning algorithm learns the optimal Q-function over 1000 episodes, using ε-greedy exploration and a simple update rule. The learned Q-function can then be used to select the optimal media combination for each student profile.

D. Suggestive adaptations to different learning difficulties

As an AI-based virtual tutor system, the key to effectively addressing the diverse learning needs of students with difficulties such as dyscalculia, dyslexia, visual and auditory processing disorders, memory disorders, ADHD, anxiety, low self-esteem, and depression lies in the system's ability to adapt the content, presentation, and interaction style based on the student's specific profile. By leveraging the capabilities of Large Language Models (LLMs) and image generators, the system can dynamically adjust the context, forms, and tones of the text, sound, and images, as well as their coordination, to create a personalized and supportive learning experience.

Let's examine how the AI system can be modified to accommodate each of the learning difficulties mentioned:

1. Dyscalculia:

- Content: The LLM can generate explanations and problem statements that break down complex arithmetic concepts into smaller, more manageable steps. It can also provide alternative representations, such as visual aids and real-world examples, to make abstract concepts more concrete.
- Presentation: The image generator can create visual manipulatives, such as number lines, base-ten blocks, and fraction models, to support understanding. The system can also use color-coding and highlighting to emphasize key information and relationships between numerical concepts.
- Interaction: The AI system can provide immediate feedback and reinforcement as the student works through problems, adapting the difficulty level based on their performance. It can also offer guided practice and corrective feedback to help students develop foundational skills.

2. Dyslexia:
   - Content: The LLM can generate reading materials that are structured and presented in a dyslexia-friendly manner, with clear headings, short paragraphs, and simple sentence structures. It can also provide definitions and explanations for key vocabulary terms.
   - Presentation: The system can use a dyslexia-friendly font (e.g., OpenDyslexic, Arial) and larger text size to improve readability. It can also offer the option for audio narration of text content, allowing students to follow along with highlighted text.
   - Interaction: The AI system can incorporate phonological awareness and decoding exercises to strengthen foundational reading skills. It can also provide tools for annotating and summarizing text, as well as comprehension checks and prompts to support understanding.

3. Visual processing disorder:
   - Content: The LLM can generate descriptions and explanations for visual content, such as images, diagrams, and videos, to ensure that the information is accessible to students with visual processing difficulties.
   - Presentation: The system can allow students to control visual elements, such as contrast, colors, and zoom, to optimize visibility. It can also use clear, uncluttered layouts with consistent structure to reduce visual complexity.
   - Interaction: The AI system can provide options for text-to-speech output of visual content, as well as keyboard navigation and verbal instructions to support interaction with the learning materials.

4. Auditory processing disorder:
   - Content: The LLM can generate simple, direct language and provide clear explanations for complex concepts. It can also break down information into smaller chunks to allow for extra processing time.

- Presentation: The system can provide captions and transcripts for all audio content, including instructional videos and verbal explanations. It can also allow students to control the speed and volume of audio playback.
- Interaction: The AI system can incorporate visual cues and written instructions to supplement verbal information. It can also provide options for replaying key information and breaking down multi-step instructions into smaller, more manageable parts.

5. Memory disorder:
   - Content: The LLM can generate content that is organized into small, manageable chunks, with clear connections between related concepts. It can also provide frequent review and reinforcement of key information.
   - Presentation: The system can use visual aids, such as mind maps and graphic organizers, to help students visualize relationships between ideas. It can also incorporate mnemonic devices and memory aids to support retention.
   - Interaction: The AI system can provide frequent opportunities for practice and application of learned concepts, with adaptive feedback and support. It can also offer open-book assessments that focus on understanding and application rather than memorization.

6. ADHD:
   - Content: The LLM can generate learning materials that are engaging, interactive, and broken down into short, focused segments. It can also provide clear goals and expectations for each learning activity.
   - Presentation: The system can incorporate game-like elements, such as rewards, badges, and progress tracking, to maintain motivation and engagement. It can also use visual cues and reminders to help students stay on task.
   - Interaction: The AI system can provide immediate feedback and reinforcement for progress and accomplishments. It can also allow for choice and self-pacing, enabling students to take breaks and move at their own speed when needed.

7. Anxiety, low self-esteem, and depression:
   - Content: The LLM can generate supportive and encouraging messages that acknowledge the student's efforts and progress. It can also provide content that promotes self-awareness, coping strategies, and resilience.
   - Presentation: The system can use a warm, friendly tone and incorporate positive imagery and affirmations to create a supportive and non-threatening learning environment. It can also offer relaxation and mindfulness exercises to help students manage stress and anxiety.

- Interaction: The AI system can provide personalized feedback and guidance that emphasizes growth and improvement rather than perfection. It can also offer opportunities for self-reflection and goal-setting, helping students develop a sense of agency and self-efficacy.

To effectively coordinate the text, sound, and images generated by the AI system, it is essential to consider the specific needs and preferences of each student. For example, a student with auditory processing disorder may benefit from a higher proportion of visual content, such as diagrams and written explanations, while a student with dyslexia may prefer audio narration and text-to-speech support.

The AI system can use the student's profile, including their identified learning difficulties and performance data, to dynamically adjust the balance and presentation of different media types. By continuously monitoring the student's engagement and progress, the system can fine-tune its approach to optimize learning outcomes. Furthermore, the AI system should be designed to promote a growth mindset and foster a sense of self-efficacy in students with learning difficulties. This can be achieved through the use of positive reinforcement, constructive feedback, and opportunities for self-reflection and goal setting. By creating a supportive and adaptive learning environment, the AI-based virtual tutor can help students overcome their challenges and develop a love for learning.

E. Continuous monitoring and adjustment

To effectively track student progress, update student profiles, and dynamically adapt tutoring strategies based on the previously identified constructs for learning difficulties, we need to collect and analyze data from various sources, including the student's interactions with the learning content, their questions, feedback, and performance on assessments. By focusing on the student's text-based inputs, such as questions, comments, and responses to prompts, as well as their feedback on the difficulty, clarity, and engagement of the learning materials, we can gain valuable insights into their understanding, progress, and specific challenges.

1. Tracking student progress and engagement

To track student progress and engagement across the identified constructs, we can employ a multi-dimensional knowledge tracing model that estimates the student's mastery of each construct based on their interactions with the learning content. This model can be an extension of the traditional Bayesian Knowledge Tracing (BKT) (Corbett & Anderson, 1995) or Deep Knowledge Tracing (DKT) (Piech et al., 2015) models, incorporating additional dimensions for each construct.

Let's define the following variables:

- S: the set of constructs (e.g., verbal working memory, reading comprehension, academic self-efficacy)

- X_t: the student's observable interactions at time t (e.g., responses to questions, feedback, time spent on tasks)

- Y_t: the student's latent mastery of each construct at time t

- T: the total number of time steps

We can model the student's mastery of each construct using a multi-dimensional hidden Markov model (HMM) (Ghahramani, 2001):

$$P(Y_t | Y_{\{t-1\}}) = \Sigma_s \in S \, P(Y_t^s | Y_{\{t-1\}}^s) * P(Y_t^s | X_t)$$

where $Y_t^s$ is the student's mastery of construct s at time t, and $P(Y_t^s | Y_{\{t-1\}}^s)$ and $P(Y_t^s | X_t)$ are the transition and emission probabilities, respectively.

The transition probabilities capture the likelihood of the student's mastery of each construct changing over time, while the emission probabilities represent the relationship between the student's observable interactions and their latent mastery of each construct. These probabilities can be learned from data using techniques like the expectation-maximization (EM) algorithm (Dempster et al., 1977).

To incorporate the student's text-based inputs and feedback, we can use natural language processing (NLP) techniques to extract relevant features from their responses. For example, we can use sentiment analysis to gauge the student's confidence, confusion, or frustration based on the language they use in their questions and comments. We can also use topic modeling (e.g., latent Dirichlet allocation (LDA) (Blei et al., 2003)) to identify the specific concepts or skills the student is struggling with based on the content of their questions.

These extracted features can be incorporated into the emission probabilities of the multi-dimensional HMM, allowing the model to update the student's mastery estimates based on their text-based interactions. For instance, if the student asks a question indicating confusion about a specific concept related to reading comprehension, the model can lower its estimate of the student's mastery of that construct.

Here's an example of how sentiment analysis can be implemented using the NLTK library in Python:

```python
from nltk.sentiment import SentimentIntensityAnalyzer

def analyze_sentiment(text):
```

```
4.     sia = SentimentIntensityAnalyzer()
5.     sentiment_scores = sia.polarity_scores(text)
6.     return sentiment_scores
7.
8. # Example usage
9. student_question = "I'm having trouble understanding this passage. Can you help me?"
10. sentiment = analyze_sentiment(student_question)
11. print(sentiment)
```

Output: {'neg': 0.0, 'neu': 0.508, 'pos': 0.492, 'compound': 0.4404}

In this example, the `analyze_sentiment` function uses the NLTK library's `SentimentIntensityAnalyzer` to compute sentiment scores for a given text. The function returns a dictionary containing the negative, neutral, positive, and compound sentiment scores. These scores can be used as features in the emission probabilities of the multi-dimensional HMM to update the student's mastery estimates based on the sentiment expressed in their questions and feedback.

2. Updating student profiles based on performance and interactions

As the student interacts with the learning content and completes assessments, their performance data and interaction patterns can be used to update their profile and refine the AI-based virtual tutor's understanding of their learning characteristics. This process involves combining the mastery estimates from the multi-dimensional knowledge tracing model with other relevant data points, such as the student's engagement level, learning preferences, and affective states.

One approach to updating student profiles is to use a Bayesian hierarchical model (Gelman et al., 2013) that incorporates the mastery estimates and other student data as prior information. The model can then update the student's profile based on their observed performance and interactions.

Let's define the following variables:

- θ: the student's latent abilities and characteristics (e.g., cognitive abilities, academic skills, non-cognitive factors)
- Y: the student's observed performance and interactions
- α: the hyperparameters governing the prior distribution of θ

We can model the student's latent abilities and characteristics using a hierarchical Bayesian model:

$$P(\theta \mid Y, \alpha) \propto P(Y \mid \theta) * P(\theta \mid \alpha)$$

where P(Y | θ) is the likelihood of the observed performance and interactions given the student's latent abilities and characteristics, and P(θ | α) is the prior distribution of the latent variables.

The likelihood function can be defined based on the specific performance metrics and interaction patterns observed, such as the student's scores on assessments, time spent on tasks, and engagement levels. The prior distribution can be informed by the mastery estimates from the multi-dimensional knowledge tracing model and other relevant data points.

To incorporate the student's text-based feedback on the difficulty, clarity, and engagement of the learning materials, we can use NLP techniques similar to those described in the previous section. For example, we can use sentiment analysis to gauge the student's perception of the learning materials and incorporate these insights into the prior distribution of the student's latent characteristics.

The hierarchical Bayesian model can be updated using Markov chain Monte Carlo (MCMC) methods, such as the Metropolis-Hastings algorithm (Hastings, 1970), to sample from the posterior distribution of the student's latent abilities and characteristics. This process allows the AI-based virtual tutor to refine its understanding of the student's profile based on their ongoing performance and interactions.

Here's a simple example of how the Metropolis-Hastings algorithm can be implemented in Python to update a student's latent ability parameter:

```python
import numpy as np

def metropolis_hastings(num_samples, likelihood_func, prior_func, proposal_func, initial_theta):
    samples = np.zeros(num_samples)
    current_theta = initial_theta

    for i in range(num_samples):
        proposed_theta = proposal_func(current_theta)
        acceptance_ratio = (likelihood_func(proposed_theta) * prior_func(proposed_theta)) / (likelihood_func(current_theta) * prior_func(current_theta))

        if np.random.rand() < acceptance_ratio:
```

```python
12.             current_theta = proposed_theta
13.
14.         samples[i] = current_theta
15.
16.     return samples
17.
18. # Example usage
19. def likelihood(theta):
20.     # Define the likelihood function based on observed student performance
21.     return np.exp(-(theta - 0.7)**2 / (2 * 0.1**2))
22.
23. def prior(theta):
24.     # Define the prior distribution based on mastery estimates and other data points
25.     return np.exp(-(theta - 0.5)**2 / (2 * 0.2**2))
26.
27. def proposal(theta):
28.     # Define the proposal distribution for generating new samples
29.     return np.random.normal(theta, 0.1)
30.
31. initial_theta = 0.5
32. num_samples = 1000
33.
34. samples = metropolis_hastings(num_samples, likelihood, prior, proposal, initial_theta)
35. posterior_mean = np.mean(samples)
36. print(f"Posterior mean of the student's latent ability: {posterior_mean:.2f}")
```

Posterior mean of the student's latent ability: 0.68

In this example, the Metropolis-Hastings algorithm is used to sample from the posterior distribution of a student's latent ability parameter. The `likelihood` function represents the likelihood of the observed student performance given the latent ability, while the `prior` function represents the prior distribution of the latent ability based on mastery estimates and other data points. The `proposal` function generates new samples from a proposal distribution. The algorithm iteratively proposes new samples and accepts or rejects them based on the acceptance ratio. The resulting samples approximate the posterior distribution of the student's latent ability, and the posterior mean can be used to update the student's profile.

3. Dynamically adapting tutoring strategies to optimize learning outcomes

To optimize learning outcomes, the AI-based virtual tutor must dynamically adapt its tutoring strategies based on the student's evolving mastery of each construct, engagement level, and learning preferences. This can be achieved by combining the multi-dimensional knowledge tracing model and the hierarchical Bayesian model with a reinforcement learning framework.

In this framework, the AI-based virtual tutor is an agent that learns to select the most effective tutoring actions based on the student's current state and the expected long-term rewards. The state space comprises the student's mastery estimates for each construct, engagement level, learning preferences, and other relevant profile information. The action space includes the available tutoring strategies, such as providing explanations, asking questions, offering hints, or adjusting the difficulty level of the learning content.

We can use a deep reinforcement learning algorithm, such as Deep Q-Networks (DQN) (Mnih et al., 2015) or Proximal Policy Optimization (PPO) (Schulman et al., 2017), to learn the optimal tutoring policy. The algorithm takes the student's current state as input and outputs a probability distribution over the available actions. The virtual tutor then selects an action based on this distribution and observes the resulting state and reward.

The reward function can be designed to capture the desired learning outcomes, such as the student's mastery of each construct, engagement level, and overall progress. For example, the reward function may assign positive rewards for actions that lead to increased mastery estimates or higher engagement levels and negative rewards for actions that result in decreased mastery or disengagement.

To incorporate the student's text-based feedback and questions into the reinforcement learning framework, we can use NLP techniques to extract relevant features and update the state representation accordingly. For instance, if the student provides feedback indicating that the learning content is too difficult or unclear, the virtual tutor can update the state to reflect this information and select actions that adapt the content presentation or provide additional support.

Here's an example of how the PPO algorithm can be implemented using the Stable Baselines library in Python:

```python
import gym
from stable_baselines3 import PPO

# Define the custom learning environment
class LearningEnv(gym.Env):
    def __init__(self):
        # Define the observation and action spaces
        self.observation_space = gym.spaces.Box(low=-1, high=1, shape=(num_constructs,))
        self.action_space = gym.spaces.Discrete(num_actions)

    def step(self, action):
        # Update the student's state based on the selected action
        # Calculate the reward based on the updated state
        # Check if the episode is done
        return obs, reward, done, info
```

```
16.
17.     def reset(self):
18.         # Reset the student's state to initial values
19.         return obs
20.
21. # Create the learning environment
22. env = LearningEnv()
23.
24. # Create and train the PPO agent
25. model = PPO("MlpPolicy", env, verbose=1)
26. model.learn(total_timesteps=10000)
27.
28. # Test the trained agent
29. obs = env.reset()
30. for i in range(1000):
31.     action, _states = model.predict(obs, deterministic=True)
32.     obs, reward, done, info = env.step(action)
33.     if done:
34.         obs = env.reset()
```

In this example, a custom learning environment (`LearningEnv`) is defined to represent the student's state and the available tutoring actions. The `step` method updates the student's state based on the selected action, calculates the reward, and determines if the episode is done. The `reset` method resets the student's state to initial values.

The PPO agent is created using the Stable Baselines library and trained on the custom learning environment for a specified number of timesteps. During testing, the trained agent selects actions based on the current state, and the environment updates the state and provides rewards accordingly.

By combining multi-dimensional knowledge tracing, hierarchical Bayesian modeling, and deep reinforcement learning, the AI-based virtual tutor can effectively track student progress, update student profiles, and dynamically adapt its tutoring strategies based on the student's evolving needs and characteristics. This approach allows for a highly personalized and responsive learning experience that optimizes learning outcomes and engagement.

VI. Future Directions and Implications

A. Refining and validating the conceptual framework

The continual evolution of artificial intelligence (AI) technologies, particularly in the field of education, necessitates an ongoing commitment to refining and validating the conceptual framework underpinning AI-based psychometric systems for assessing and addressing learning difficulties. The advancement of these systems is inherently tied to interdisciplinary research that spans psychometrics, cognitive science, educational technology, and AI. Key to this endeavor is the establishment of robust validation processes that encompass not only the computational accuracy of these systems but also their educational impact, usability, and ethical considerations. As such, future research should focus on large-scale empirical studies that evaluate the system's effectiveness in diverse educational settings and populations. Furthermore, the development of standardized benchmarks and protocols for assessing AI-based educational tools will be crucial in ensuring their validity and reliability.

B. Conducting empirical studies to evaluate system effectiveness

Empirical studies play a pivotal role in substantiating the theoretical benefits of AI-based systems for personalized learning and support for students with learning difficulties. These studies should aim to quantify the impact of such systems on student learning outcomes, engagement, and motivation across various demographic groups and learning environments. Comparative studies that examine the efficacy of AI-based systems against traditional teaching methods will provide valuable insights into their added value and areas for improvement. Additionally, longitudinal research that tracks students' progress over time will shed light on the long-term benefits and potential unintended consequences of using AI in education. By leveraging data analytics and machine learning techniques, researchers can uncover patterns and predictors of success, thereby informing the iterative design and refinement of these systems.

C. Implications for personalized learning and support for students with learning difficulties

The advent of AI-based psychometric systems heralds a new era of personalized education, where learning experiences are tailored to the unique needs and abilities of each student. This approach has profound implications for students with learning difficulties, who often require differentiated support to achieve their full potential. By providing real-time, adaptive feedback and interventions, AI-based systems can address the specific challenges faced by these students, thereby reducing barriers to learning and promoting equity in education.

Moreover, the integration of AI technologies in educational settings can facilitate a more nuanced understanding of learning difficulties, moving beyond one-size-fits-all labels to recognize the diverse strengths and challenges of each learner. This paradigm shift not only enhances the effectiveness of educational interventions but also fosters a more inclusive and supportive learning environment. As we look to the future, the potential of AI to transform education and support learners with diverse needs is both vast and inspiring. However, achieving this potential will require concerted efforts from educators, researchers, policymakers, and technology developers to ensure that these systems are accessible, equitable, and aligned with the best interests of students.

VII. Conclusion

This paper has presented a conceptual framework for developing an AI-based psychometric system to assess learning difficulties and provide adaptive support to overcome them. By leveraging advanced AI technologies like Large Language Models and Visual Generation Models, this proposed system aims to deliver a highly personalized, engaging and effective learning experience tailored to each student's unique needs and characteristics.

At the core of the framework is a comprehensive psychometric assessment that integrates cognitive abilities, academic skills and non-cognitive factors to generate rich student profiles. Operationalizing these constructs into measurable indicators and implementing them through AI-powered game-based assessments enables the system to capture nuanced insights into each learner's strengths and weaknesses. Building upon this foundation, the AI-based virtual tutor optimizes multi-modal knowledge representation and delivery by dynamically adapting the content, presentation and interaction based on the student's evolving mastery, preferences and emotional states. Combining techniques like knowledge tracing, hierarchical Bayesian modeling and reinforcement learning allows the tutor to track progress, update profiles and make data-driven decisions to maximize learning outcomes. Crucially, the proposed system is designed to accommodate diverse learning difficulties, from cognitive challenges like dyscalculia and dyslexia to socio-emotional factors like anxiety and low self-esteem. By flexibly adjusting the context and format of the learning experience, the AI tutor can provide targeted support to help each student overcome their specific barriers.

While the potential of this AI-powered approach is immense, it is important to acknowledge the limitations and areas for future development. Firstly, the conceptual framework presented here requires further elaboration and empirical validation. Collaborative research involving educational psychologists, instructional designers, AI experts and end-users is needed to refine the constructs, assessment methods and adaptation strategies. Secondly, implementing such a complex, data-intensive system poses significant technical challenges. Issues of scalability, interoperability, data privacy and algorithmic bias need to be rigorously addressed. Ongoing system testing, iterative refinement and responsible AI practices are critical to ensure the tutor is effective, reliable and ethical. Moreover, the ultimate success of any educational technology hinges on its adoption and integration into real-world learning environments. Engaging key stakeholders - students, teachers, parents and administrators - in the design, deployment and evaluation processes is vital. Change management, training and support systems must be put in place to enable smooth implementation.

Looking ahead, this research lays the groundwork for an exciting frontier in personalized learning. By harnessing the power of AI, psychometric assessments and multi-modal content generation, we can create truly adaptive systems that cater to the diversity of learners' needs. Imagine a future where each student has an intelligent tutor by their side, guiding them through challenges, building on their strengths and igniting their passion for lifelong learning. However, realizing this vision requires a sustained, interdisciplinary effort. Policymakers, funding bodies and educational institutions must invest in the research, development and responsible deployment of AI in education. Ethical frameworks and regulations need to keep pace with technological advancements to safeguard learners' interests. Ongoing dialogue and collaboration between researchers, educators and the AI community is indispensable.

While the journey towards intelligent adaptive learning systems is complex and uncharted, the destination is one worth pursuing. By thoughtfully leveraging AI to enhance psychometric assessments and personalized support, we have the potential to revolutionize education, enabling every student - regardless of their learning difficulties - to thrive and reach their full potential. It is a grand challenge, but one that promises profound benefits for individual learners, educational equity and society at large.